# Flexible manipulation of chiral spin state by chemical bond in Mn triangular lattice magnet

Jiyuan Xu, Xin Liu, Li Ma*, Guoke Li*, Dewei Zhao, Congmian Zhen, and Denglu Hou

*Hebei Key Laboratory of Photophysics Research and Application, College of Physics, Hebei Normal University, Shijiazhuang, 050024, China.*

**Abstract:**

This study investigates the influence of chemical bonds on the magnetic structure of materials, a less explored area compared to their effect on crystal stability. By analyzing the strength and directionality of chemical bonds using the electron localization function (ELF) and charge density difference (CDD) methods, we examine their impact on magnetic exchange interactions and magnetocrystalline anisotropy under specific interstitial conditions in $Mn_4X$ compounds. Our findings indicate that these properties can effectively modulate the magnetic ground state. This work not only elucidates the varied magnetism observed in Mn triangular lattice magnets but also proposes an approach for engineering chiral spin states through chemical bonding manipulation.

## I. INTRODUCTION

The triangular lattice magnets can exhibit a rich variety of spin states due to different sublattice exchange interactions, [1-5] including one-dimensional collinear, two-dimensional non-collinear, and three-dimensional non-coplanar spin states. Among these, the triangular lattice magnets with the non-coplanar spin states have been theoretically expected to exhibit many emerging physical phases and effects. The magnetic skyrmion phase [6-8] has been predicted in the triangular lattice magnets

* Corresponding author. E-mail address: majimei@126.com.
* Corresponding author. E-mail address: liguoke@126.com.

NiGa$_2$S$_4$ [6], NiBr$_2$ [6] and Pr(Ti,V,Tr)$_2$(Al,Zn)$_{20}$ [8]. It is also a good platform for cultivating the spin liquid state [9]. Besides, topological magnetoelectric transport effects involving the anomalous Hall effect [10], the spontaneous quantum Hall effect [11,12] and the quantum topological Hall effect [13] have been predicted in the noncoplanar triangular lattice antiferromagnets. Recently, topological magneto-optical effects have also been predicted in the noncoplanar triangular lattice antiferromagnets [14]. These physical effects related to the chiral spin states have also been discovered in experiments. An unconventional anomalous Hall effect is revealed in the quasi-two-dimensional triangular lattice antiferromagnet PdCrO$_2$ below 20 K [15]. This effect is also observed in the quasi-two dimensional triangular lattice helimagnet Fe$_{1.3}$Sb below 30 K [16]. Besides, the geometric Hall effect in the noncoplanar triangular lattice antiferromagnet UCu$_5$ below 15 K has emerged [17]. skyrmion lattice [18,19] and topological Nernst effect [19] has been observed in the frustrated centrosymmetric triangular lattice magnet Gd$_2$PdSi$_3$ below 20 K. Recently, spontaneous topological Hall effect has been verified experimentally in the noncoplanar triangular lattice antiferromagnets CoTa$_3$S$_6$ and CoNb$_3$S$_6$ below 20 K [20].

All of the above discoveries originate from noncoplanar chiral spin states in the triangular lattice. However, the temperature at which physical effects are realized is generally low. Therefore, how to construct stable chiral spin states becomes extremely important. In this case, understanding and manipulating the internal driving forces for the formation of chiral spin state in triangular lattice is crucial. At present, theoretical research has proposed several schemes for constructing chiral spin states, such as spin-charge coupling between itinerant electron spins and localized spins, [3,4,7,11,12,21] frustrated exchange interactions and dipole interactions in localized spin systems, [6,22,23] and the interplay between the spin-orbit coupling and itinerant magnetism [24]. Lately, it is predicted that the chiral spin state can also be stabilized by the RKKY interactions within $f^2$ non-Kramers doublet systems [8]. All of the above mechanisms only involve the physical level of electrons and not the chemical level of electrons. Electrons are the binding agents of materials. Figure. 1 shows the three functions of the $d$ electrons, namely providing exchange interaction, forming localized spin and chemical bonds. Among them, it is known that the *d-d* exchange interaction determines the order temperature of the system, which together with the localized spin determines the spin state of the system, while the chemical bonds of *d* electrons determine the stability of the system. At the same time, the three functions influence each other. On the one hand, the chemical bonding of *d* electrons will lead to the redistribution of *d* electron clouds, which can change the exchange interaction. On the other hand, the chemical bonding of *d* electrons will also cause it to be unable to form a localized spin. In the present study, we push forward these theoretical studies to a more realistic situation, by taking into account both the chemical bonding and the localized spins in triangular lattice magnet. Our aim is to illuminate chiral spin state induced by the interplay between the two couplings. To this end, we choose the Mn triangular lattice as our research platform due to the fact that Mn atoms have half-filled *d* orbitals which can provide a large localized spin and easily form chemical bonds with surrounding elements.

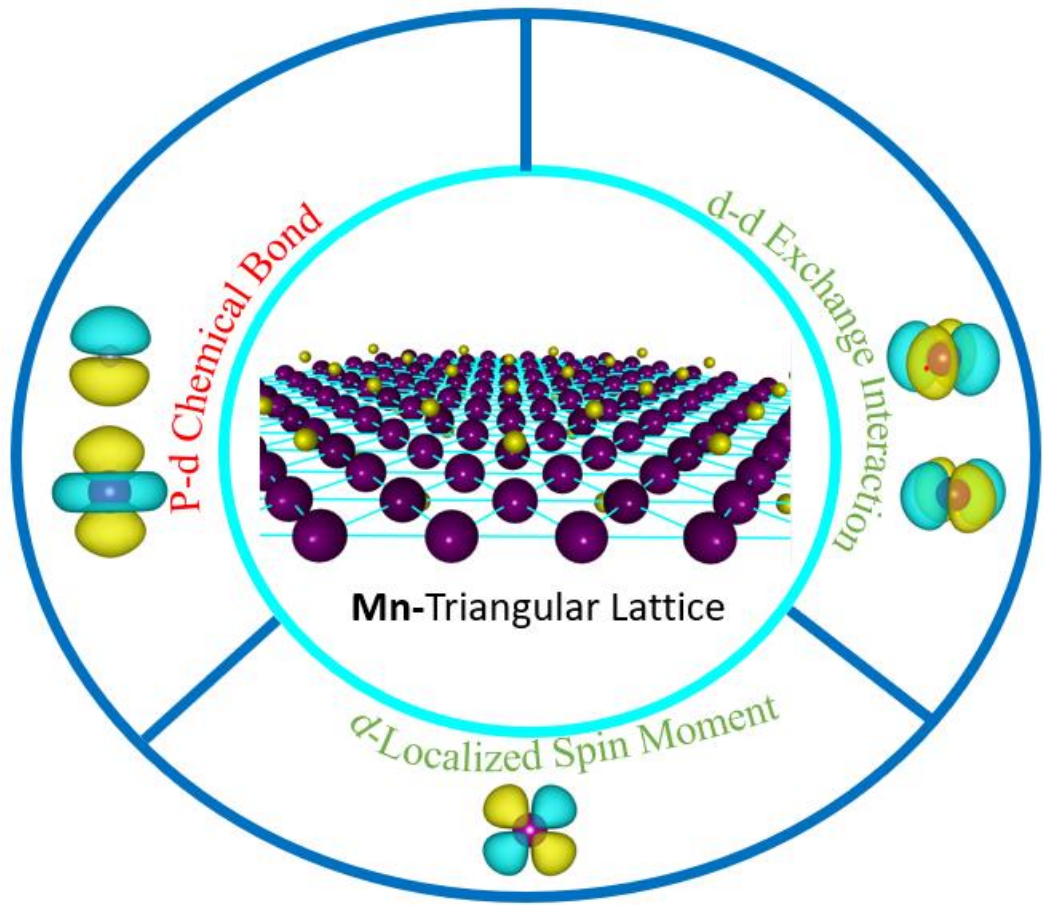

Figure. 1. Three functions of the outer *d* electrons in the Mn triangular lattice magnet, providing

exchange interaction, forming localized spin moments and chemical bonds. They are both a whole and restrict each other, thus determining the stability, magnetism and spin state of the system.

This paper is organized as follows: The methods to perform the first-principles calculation are presented in Sec. II. The magnetic structure of the ground state under different interstitials is calculated and analyzed in Sec. III A. The types and strengths of chemical bonds under different interstitials are calculated and analyzed in Sec. III B. In Sec. III C, the relationship between chemical bonds and magnetic structures is discussed. Finally, Sec. IV. contains a summary of this work.

## II. Methods

All first-principles calculations were carried out within the framework of density functional theory (DFT) and implemented using the projector-augmented wave potentials in the Vienna Ab-initio Simulation Package (VASP). Integrations were performed over a 12 ×12×12 $k$-point mesh in the Brillouin zone, and a plane wave cutoff energy of 510 eV was employed. Based on the known crystal structure, the PBE (Perdew-Burke-Ernzerhof) potential and spin-orbit coupling (SOC) were used to compare the energies among different spin configurations as well as electronic properties and exchange integral constants. We chose a lattice constant of 3.85 Å at 0 K [25] for the system and fixed the lattice constant for calculation. The electronic structure was converged to $10^{-6}$ eV/cell. In the classical Heisenberg spin system, considering the specific exchange interaction $J_{12}$ between spin sites 1 and 2, the spin Hamiltonian can be expressed as [26]

$$E_{\text{spin}} = J_{12}\mathbf{S}_1 \cdot \mathbf{S}_2 + \mathbf{S}_1 \cdot \mathbf{K}_1 + \mathbf{S}_2 \cdot \mathbf{K}_2 + E_{\text{other}},$$

Where $\mathbf{K}_1 = \sum_{i \neq 1,2} J_{1i}\mathbf{S}_i$, $\mathbf{K}_2 = \sum_{i \neq 1,2} J_{2i}\mathbf{S}_i$, $E_{\text{other}} = \sum_{i,j \neq 1,2} J_{ij}\mathbf{S}_i \cdot \mathbf{S}_j$. Here, $\mathbf{K}_1, \mathbf{K}_2$ and $E_{\text{other}}$ do not depend on the spin directions of sites 1 and 2. Four collinear spin states:

(1) $S_1 = S, S_2 = S$, (2) $S_1 = S, S_2 = -S$, (3) $S_1 = -S, S_2 = S$, (4) $S_1 = -S, S_2 = -S$ were considered. These four states have the following energy expression:

$$E_1 = E_0 + E_{other} + J_{12}S^2 + K_1S + K_2S$$
$$E_2 = E_0 + E_{other} - J_{12}S^2 + K_1S - K_2S$$
$$E_3 = E_0 + E_{other} - J_{12}S^2 - K_1S + K_2S$$
$$E_4 = E_0 + E_{other} + J_{12}S^2 - K_1S - K_2S$$

Then $J_{12}$ can be written as

$$J_{12} = \frac{E_1 + E_4 - E_2 - E_3}{4S_1S_2}$$

and the exchange integral $J_{12}$, which can determine the interaction between ions, will be obtained.

## III. RESULTS AND DISCUSSION

### A. Magnetic Structure

Figure. 2 (a) shows that the two-dimensional (2D) Mn triangular lattice magnet is composed of 4 magnetic sublattices. Our study focuses on exploring the relationship between chemical bonds and magnetic structures. It is found that the outer *d* electrons of transition elements can form different bonds with different small interstitial atoms [27]. Besides, these small interstitial atoms can also effectively improve the corrosion resistance, wear resistance and stability of the system [28-31]. In this work, we intend to modify the *d*-orbital bonding of Mn by different interstitial settings in adjacent Mn triangular lattices. The specific implementation scheme is shown in Figure. 2 (b). Here, the three-dimensional (3D) Mn triangular lattice magnet with ABC close stacking is adopted, where the (111) lattice plane of 3D Mn triangular lattice circled by the blue dotted line corresponds to the 2D Mn triangular lattice in Figure. 2 (a), and the yellow solid circle represents different interstitial settings including vacancies, N atoms, and C atoms, which are inserted into the nearest-neighbor Mn triangular lattice planes. Mn triangular lattice under the vacancy, N, and C interstitials correspond to the real materials *γ*-Mn, $Mn_4N$, and $Mn_4C$, respectively. This ensures that our calculation results can be compared and analyzed with the theoretical and experimental results of the above real materials. The spin state of the triangular lattice can be represented by the angle $\theta$ between the sublattice spins [3,14,32], as shown in Figure. 2 (c). Different $\theta$ corresponds to different spin states. $\theta$ =70.5° corresponds to 2-in 2-out chiral spin state, $\theta$ =109.5° corresponds to all-in all-out 3-q chiral spin state, when $\theta$ is equal to 0°, it is a 3:1 collinear 1-q spin state, etc [3,14,32]. Therefore, an energy scan calculation from 0° to 180° is performed on $\theta$ to find the magnetic ground state under different interstitial settings. The calculation results are shown in Fig. 2 (d-f).

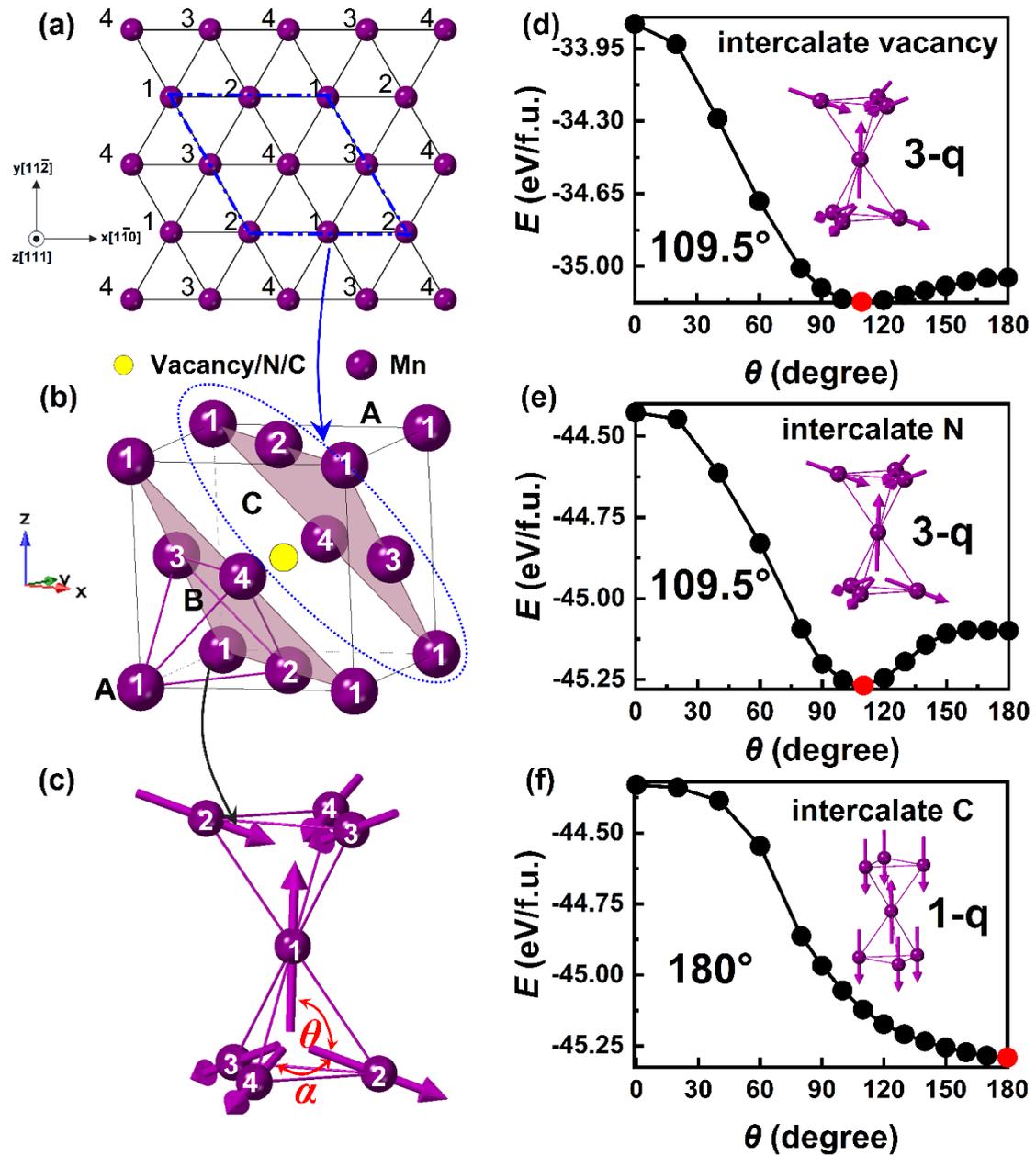

Figure. 2. Spin states of the ABC-stacking Mn triangular lattice magnet under vacancy interstitial, N interstitial, and C interstitial. (a) 2D Mn triangular lattice, it is the (111) cross section of a 3D-

ABC stacking Mn lattice, the numbers 1–4 label the four magnetic sublattices. (b) 3D-ABC stacking of Mn crystal structure, the purple spheres are Mn atoms, and the yellow solid circles are interstitial positions that can be occupied by vacancies, N, or C. (c) Magnetic structure unit in the ABC-stacking Mn triangular lattice magnet with four sublattices, and the angle $\theta$ between the spin direction of the sublattice 2 and the spin direction of the sublattice 1. Energy scans of the angle $\theta$ from 0° to 180° in the ABC-stacking Mn triangular lattice magnet under vacancy interstitial (d), N interstitial (e), and C interstitial (f). Keep the spin direction of Mn1a unchanged and gradually change the spin direction of Mn3c, the insets are the corresponding most stable spin state.

Figure. 2 (d) shows that the Mn triangular lattice under the vacancy interstitial has a magnetic ground state at $\theta = 109.5°$, corresponding to an all-in all-out 3-q chiral spin state as shown in the inset of Fig. 2 (d). This kind of spin state has been predicted in the 2D monolayer Mn triangular lattice [32]. Our calculation results demonstrates that the 3D Mn triangular lattice with ABC stacking does not destroy its chiral spin state. However, it is found that the energy of the system in Fig. 2 (d) is ~ 10 eV/f. u. higher than that in Fig. 2 (e) and Fig. 2 (f), so this system is not stable in reality. Due to this reason, $\gamma$-Mn only exists between 1095 °C and 1134 °C [33]. Furthermore, our calculations reveal that the interstitial of small *sp* elements such as N and C can indeed greatly improve the stability of the system, which may also be the reason why $Mn_4N$ and $Mn_4C$ can exist stably in nature.

In the case of the N interstitial in Fig. 2 (e), the Mn triangular lattice still reaches the magnetic ground state at $\theta = 109.5°$, i.e., the all-in all-out 3-q chiral spin state as shown in the inset of Fig. 2 (e). This is in striking agreement with the experimental results obtained by Fruchart *et al.* using neutron diffraction [29,34]. Furthermore, *Ab*-initio calculations on $Mn_4N$ by Uhl *et al* [35]. and Zhang *et al* [36]. also predicted this magnetic ground state. In addition, the energy difference $\Delta E$ between the 3-q magnetic ground state at $\theta = 109.5°$ and the 1-q magnetic excited state at $\theta = 180°$ is calculated. The $\Delta E$ value is 0.12 eV/*f.u.* and 0.17 eV/f.u. for $\gamma$-Mn and $Mn_4N$, respectively, which indicate that the N interstitial of the Mn triangular lattice not only does not destroy the chiral spin state but also makes it more stable. Therefore, the stable Mn$_4$N is expected to exhibit physical phases and effects originating from its robust chiral spin states.

As can be seen from Figure. 2 (f), the system in the C interstitial setting does not reach a minimum in the 3-q state as in the first two cases, but instead reaches a minimum at $\theta$ = 180°, corresponding to the 3:1 collinear ferrimagnetic 1-q state as shown in the inset of Figure. 2 (f). Theoretically, this spin state has been predicted in a triangular lattice with nearest-neighbor antiferromagnetic (AFM) coupling [1,3,37]. Experimentally, although Mn$_4$C is unstable and difficult to prepare, Si *et al.* still obtained the Mn$_4$C phase and observed its thermally enhanced magnetism [38]. Subsequently, Wang *et al.* prepared a high-purity Mn$_4$C phase and confirmed its magnetic structure by neutron diffraction, which is exactly the same as our calculated results [25]. The $\Delta E$ between the 1-q magnetic ground state and the 3-q magnetic excited state in Mn$_4$C is 0.17 eV/*f.u.*, which is exactly the opposite of the $\Delta E$ in $\gamma$-Mn and Mn$_4$N.

Our calculation results above demonstrate that the Mn triangular lattice magnet can achieve different magnetic ground states under different interstitial settings, which suggests that the magnetic ground state transition between 1-q and 3-q may be achieved through interstitial substitution. Next, from the perspective of magnetic exchange interaction energy, we analyze why different interstitial settings in Mn triangular lattice magnets can strongly affect the magnetic ground state.

In systems with localized moments the magnetic exchange interaction energy can be described by the Heisenberg model $\sum_{i \neq j} J_{ij}(\boldsymbol{S}_i \cdot \boldsymbol{S}_j)$, where $J_{ij}$ is the exchange integral constant between sites *i* and *j*, $\boldsymbol{S}_i$ and $\boldsymbol{S}_j$ are the spin moment on the sites *i* and *j*, respectively. In this model the spin moments $\boldsymbol{S}_i = S\boldsymbol{e}_i$ are well saturated[39] and thus do not vary in their magnitude $S$ for different directions $\boldsymbol{e}_i$. In this case, the magnetic exchange interaction energy is completely determined by $J_{ij}$. In contrast, in systems which show itinerant electron magnetism, the Stoner model is often employed, thus the magnitude of the spin moments $S$ depend sensitively on the splitting of the electronic band configuration. In this scenario, the magnitude of $S$ reflecting the degree of electron itinerancy also determines magnetic exchange interaction energy. Considering the localized and itinerant nature of 3$d$ electrons in transition elements, especially for Mn, the exchange integral constant $J_{ij}$ and the magnitude of the spin moment $S$ are calculated. Because $J_{ij}$ decays rapidly with increasing distance, only the $J_{ij}$ between the nearest neighboring magnetic atoms is considered here.

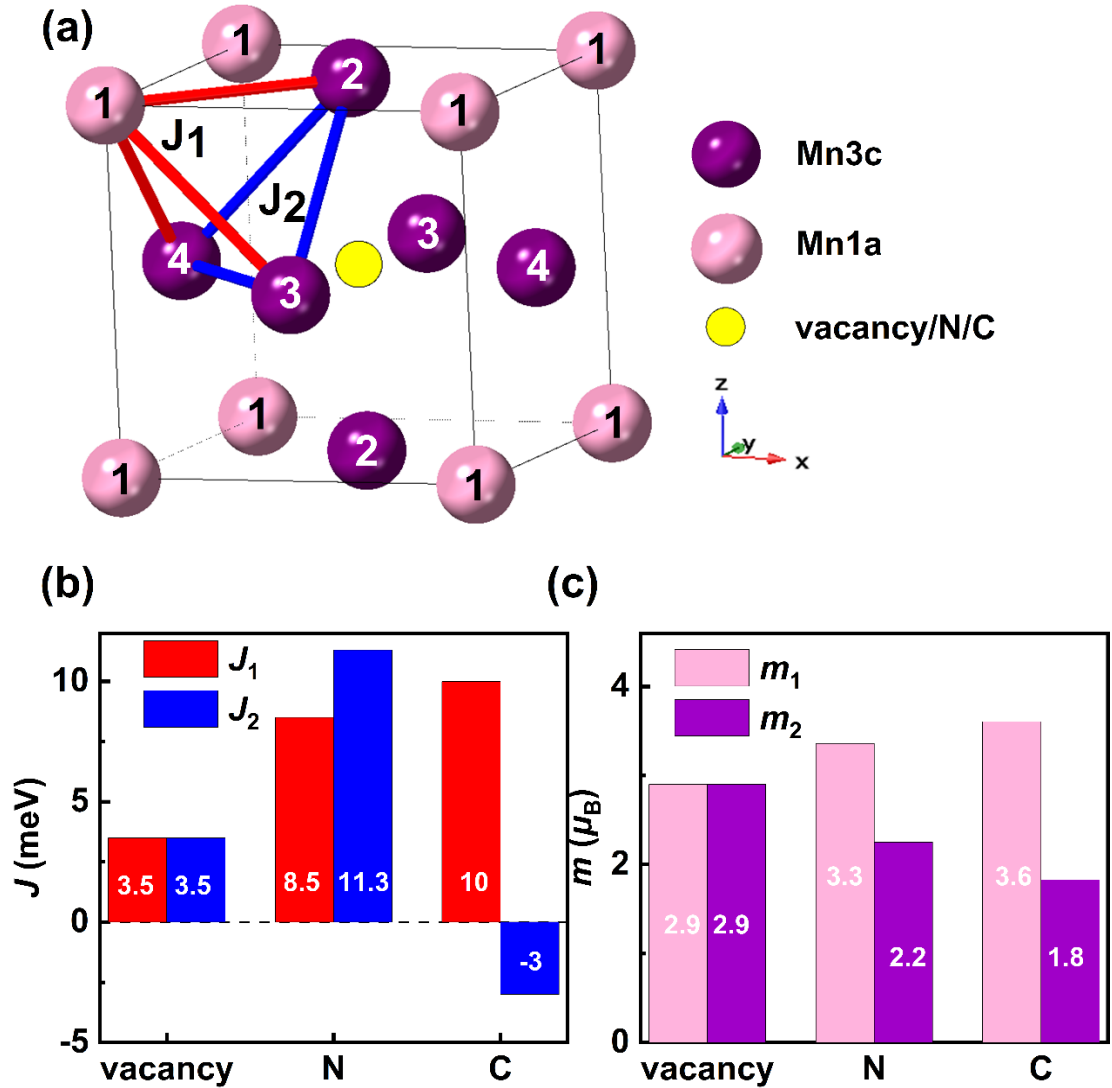

Figure. 3. Exchange integral constant $J$ and spin moment $m$ in $\gamma$-Mn, Mn$_4$N and Mn$_4$C under the magnetic ground state. (a) Mn$_4$X crystal structure. The pink spheres are Mn1a atoms occupying sublattice 1, the purple spheres are Mn3c atoms occupying sublattices 2, 3, and 4, and the yellow

solid circles are X atoms representing interstitial positions, which can be occupied by vacancies, N or C. $J_1$ is the nearest neighbor exchange integral constant between Mn1a and Mn3c, and $J_2$ is the nearest neighbor exchange integral constant between Mn3c and Mn3c. Interstitial position dependence of the exchange integral constant $J$ (b) and the spin moment $m$ (c) in γ-Mn, Mn$_4$N and Mn$_4$C.

To facilitate the analysis of the magnetic exchange interaction energy, Figure. 3 (a) shows the crystal structure of Mn$_4$X, where X represents vacancies, N or C, corresponding to γ-Mn, Mn$_4$N and Mn$_4$C, respectively. In γ-Mn, the symmetry of the sublattices 1, 2, 3 and 4 of Mn atoms is the same, while in Mn$_4$N and Mn$_4$C the symmetry is broken, leaving one Mn1a site corresponding to sublattice 1 and three equivalent Mn3c sites corresponding to sublattices 2, 3 and 4, thus compared to γ-Mn we have to take into account, that because of the two different manganese sites, Mn1a and Mn3c, additional exchange integral constants and spin moments are necessary. In Mn$_4$N and Mn$_4$C, there are two different nearest-neighbor exchange integral constants, $J_1$ between the Mn1a site and the Mn3c site and $J_2$ between the two Mn3c sites, and two spin moments, $m_1$ for Mn atoms on Mn1a sites and $m_2$ for Mn atoms on Mn3c sites. In this way, the magnetic interaction energy $E_m$ in a unit cell Mn$_4$X can be uniformly expressed as

$$E_m = 12\, J_1\, m_1\, m_2 \cos\theta + 12\, J_2\, m_2\, m_2 \cos\alpha \quad (1)$$

Where positive values of $J_1$ and $J_2$ denote AFM coupling, and the definition of angles $\theta$ and $\alpha$ is shown in Figure. 2 (c).

Figure. 3 (b) shows the interstitial position dependence of $J_1$ and $J_2$. The influence of interstitial atoms on $J_1$ and $J_2$ is analyzed from the aspects of sign and magnitude. It can be seen that $J_1$ is always positive, which means that the localized moments on Mn1a and Mn3c sites are always in AFM coupling regardless of the interstitial setting. The interstitial setting only affects the magnitude of $J_1$, which is the smallest under vacancies, dramatically increases under N interstitials, and then slightly increases under C interstitials, indicating that the interstitial setting from vacancy to N and then to C strongly enhances the electron cloud overlap between the Mn1a site and the Mn3c site.

This may be the reason why the highest Neel temperature in Mn4N and Mn4C are as high as 750 K [28-31] and 870 K [25,38,40,41], respectively.

For the case of $J_2$, it is found that both its sign and magnitude are strongly affected by the interstitial settings. Under vacancy setting, $J_2$ is the same as $J_1$ due to crystal symmetry. The $J_2$ value (11.3 meV) under N interstitials is more than three times larger than that (3.5 meV) under the vacancy interstitials, reflecting the enhanced AFM coupling of the localized moments on the two Mn3c sites, while the $J_2$ value (-3 meV) under C interstitials is negative, indicating that the coupling type of the localized moments on the two Mn3c sites has been converted to ferromagnetic.

In view of the fact that the distance between the interstitial site and the Mn1a site $\frac{\sqrt{3}a}{2}$ is nearly twice as large as that between the interstitial site and the Mn3c site $\frac{a}{2}$ ($a$ is the lattice constant of Mn4X), the sharp change of $J_2$ compared with $J_1$ should be related to some kind of close-range direct interaction introduced between the electrons from Mn3c and those from the interstitial.

Figure. 3 (c) shows the interstitial position dependence of $m_1$ and $m_2$. It is seen that the interstitial setting from vacancy to N and then to C makes the magnetic moment of the Mn1a site $m_1$ increase slowly from 2.9 $\mu_B$ to 3.3 $\mu_B$ and then to 3.6 $\mu_B$, getting closer and closer to the 5 $\mu_B$ of isolated Mn atoms, indicating that the localization of 3$d$ electrons at the Mn1a site is getting stronger and stronger. In contrast to the above situation, the magnetic moment of Mn3c position $m_2$ decreases sharply from 2.9 $\mu_B$ to 2.2 $\mu_B$ and then to 1.8 $\mu_B$ which is only half of $m_1$, indicating that the itinerancy of 3$d$ electrons at the Mn3c site or the close-range direct interaction introduced between the 3$d$ electrons from Mn3c and the 2$p$ electrons from the interstitial is getting stronger and stronger.

The magnetic ground state configuration of Mn4X is determined by the magnetic exchange interaction energy $E_m$, which is in turn determined by $J_1$, $J_2$, $m_1$ and $m_2$. The above analysis shows that the close-range direct interactions strongly affect their values, especially the values of $J_2$ and $m_2$. Chemical bonds are typical close-range direct interactions. Therefore, we will try to clarify the factors that affect $J_1$, $J_2$, $m_1$ and $m_2$ through chemical bonds. To this end, we first analyze the chemical bonds in $Mn_4X$.

## B．Chemical bonds under different interstitials

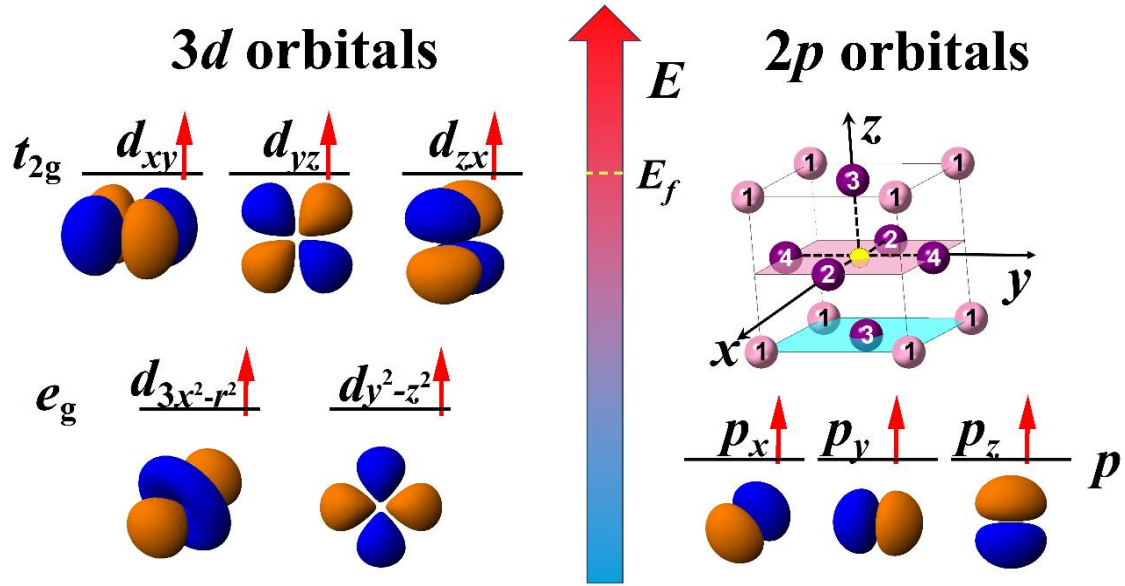

Figure. 4. Schematic diagram of the energy of 3*d* and 2*p* electrons and their orbital shape in a cubic crystal field. In $Mn_4X$, the 3*d* electrons come from Mn occupying Mn1a (pink spheres) and Mn3c (purple spheres) sites and the 2*p* electrons come from X occupying interstitial (yellow solid circle) sites. In cubic crystal field, the 3*d* orbitals involve the $d_{3x^2-r^2}$ and $d_{y^2-z^2}$ orbitals forming the lower energy 2-fold $e_g$ orbitals, and the $d_{xy}$, $d_{yz}$ and $d_{zx}$ orbitals forming the higher energy 3-fold $t_{2g}$ states. $E$ and $E_f$ represent energy and the Fermi level, respectively. The red arrows represent the electrons in the orbitals and their spins.

The type and strength of chemical bonds between electrons are determined by the energy of the electrons and the shape of their orbits, which in a crystal are strongly influenced by the crystal field and the geometric positions of atoms resulting from the crystal symmetry. The system in this work is $Mn_4X$, so the electrons involved in bonding include the outer 3*d* and 4*s* electrons of the Mn atoms and the outer 2*s* and 2*p* electrons of the interstitial X atoms. Since the 4*s* electrons of Mn are very itinerant and the 2*s* electrons of X are already paired, they cannot form bonds, so only the 3*d* of Mn and the 2*p* of X are considered here, which are in the cubic crystal field due to the ABC stacking of the Mn triangular lattice. Figure. 4 is a schematic diagram of the energy of

3d and 2p electrons and their orbital shape in a cubic crystal field. Based on our density of states calculations not given here and the results given by Goodenough for $\gamma$-Mn [42] the 5-fold degenerated 3d energy levels of isolated Mn atoms are lifted by the cubic crystal field, and the lower energy 2-fold $e_g$ states including $d_{3x^2-r^2}$ and $d_{y^2-z^2}$ orbitals and the higher energy 3-fold $t_{2g}$ states including $d_{xy}$, $d_{yz}$ and $d_{zx}$ orbitals where the Fermi level $E_f$ is located are formed.

There are three main types of chemical bonds: metallic bonds, covalent bonds and ionic bonds. Both metallic bonds and covalent bonds are shared electron interactions, but in metallic bonds, the bonding electrons are delocalized to form a "sea" of electrons, and the valence bonds are isotropic, so they do not contribute to the localized spin moment and exchange integral. In contrast, in covalent bonds, the shared electron pairs are perfectly localized between the two atoms, and have strong anisotropy in the overlapping direction of the electron cloud, *i.e.*, the orbital, thus contributing to the localized spin moment and exchange integral. Ionic bonds are formed after a complete transfer of bonding electrons from one atom to another, rarely have any particular directionality, thus contributing to the localized spin moment or even to the exchange integral due to changes in the degree of orbital filling. Therefore, below we will focus more on covalent bonds and ionic bonds.

The types of chemical bonds can be distinguished by the electron localization functions (ELF). Reference [43] analyzes the characteristics of different chemical bonds on ELF: (i) the most characteristic feature of ELF for metallic bonding is the smoothness and that it is almost constant in the interaction regions, (ii) a common ELF characteristic of all covalent bonds are the non-nuclear maxima which are present in the interaction region, (iii) ionic bonds are characterized by very low ELF values (less than ~ 0.01) in the region between atomic nuclei due to the lack of shared electrons.

Figure. 5 (a-c) shows the ELF mapping on the pink (001) lattice plane shown in Figure. 4, and Figure. 5 (d-f) shows the ELF mapping on the blue (001) lattice plane shown in Figure. 4. From left to right are the cases of vacancy interstitials, N interstitials, and C interstitials respectively. It is worth mentioning that the pink (001) lattice plane shown in Figure. 4 only contains Mn atoms on the Mn3c site and interstitial atoms on the X site, while the blue (001) lattice plane shown in Figure. 4 only contains Mn atoms on the Mn1a site and Mn3c site. The black contour lines from 0.3 to 0.5 representing metallic bonds for Fig. 5 (a) (d) (e) and (f) and from 0.7 to 1.0 representing covalent bonds for Fig. 5 (b) and (c) are drawn.

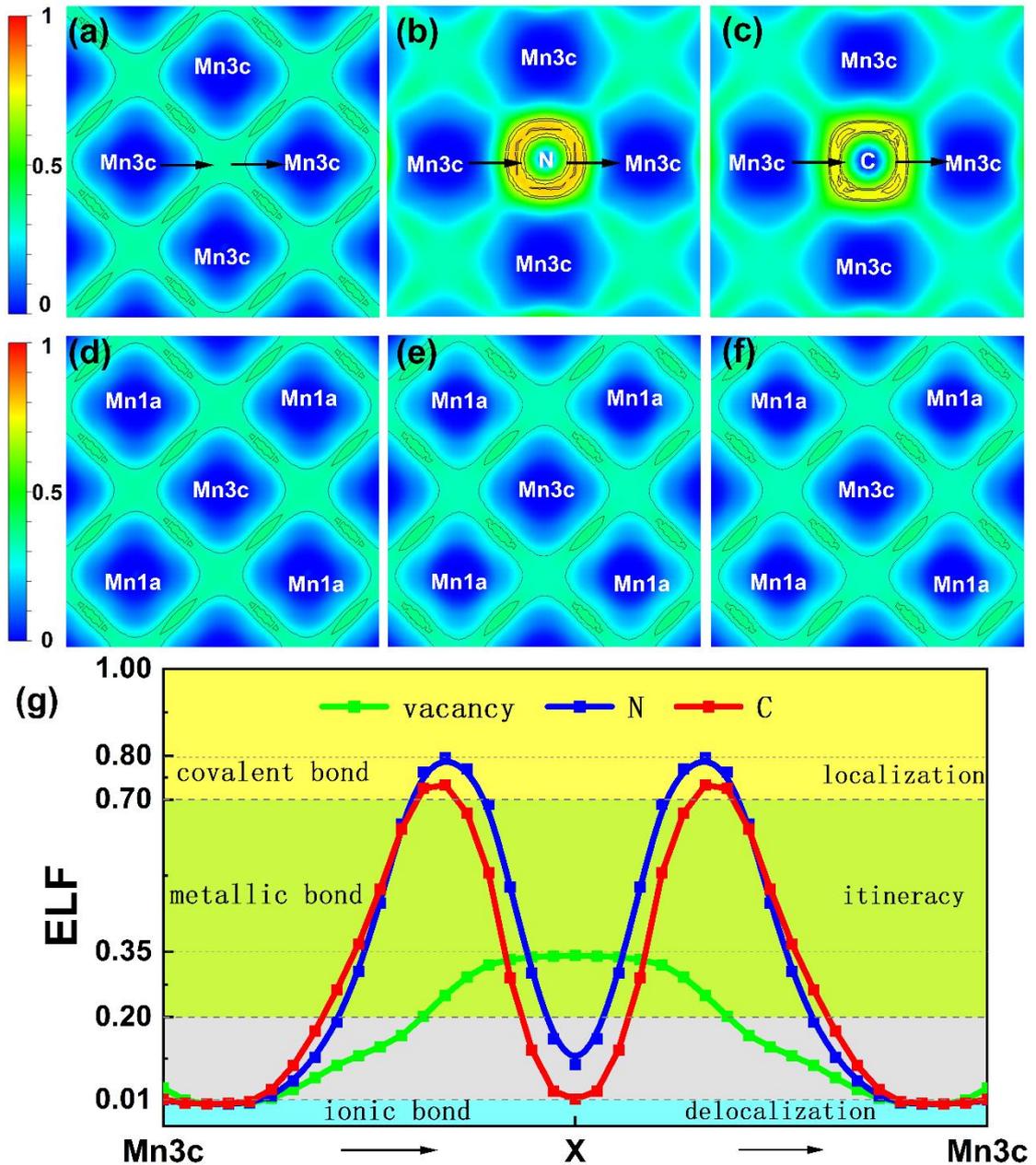

Figure. 5. Types and spatial distribution of chemical bond in $Mn_4X$. Electron localization function (ELF) mapping on the pink (001) lattice plane shown in Figure. 4 under vacancy interstitial (a), N interstitial (b), and C interstitial (c). ELF mapping on the blue (001) lattice plane shown in Figure. 4 under vacancy interstitial (d), N interstitial (e), and C interstitial (f). The contour lines in (a), (d),

(e) and (f) are from 0.3 to 0.5, and the contour lines in (b) and (c) are from 0.7 to 1.0. (g) ELF profiles selecting the path from the Mn3C site to the X site and then to the Mn3c site as indicated by the black arrows in (a-c).

There are three characteristics on the value and distribution of ELF in Fig. 5 (a-f). First, the values of ELF between the Mn1a and Mn3c sites under all different interstitial settings in Fig. 5 (d-f) and between the Mn3c sites under the vacancy interstitial in Fig. 5 (a) are all around 0.4, and the distribution of ELF is isotropic and uniform. The above characteristics are consistent with metallic bonds. Besides, a weak anisotropy distribution of ELF between Mn atoms and its value slightly higher than 0.4 also appear, as shown by the black elliptical contours in Figure. 5 (a, d-f), indicating that there may be weak covalent bond between Mn atoms except metallic bonds. The above results show that whether interstitial or not does not change the chemical bond type between Mn atoms. Second, under N interstitial, in addition to the metallic bonds between Mn atoms, as shown by the black contour lines in Figure. 5 (b), the ELF value between Mn atoms and N atoms close to the N atom side suddenly increases to above 0.7, and its distribution is a circular ring, indicating that a polar covalent bond is formed between the Mn atom and the N atom. Third, under C interstitial, as shown by the black contour lines in Figure. 5 (c), the ELF value between Mn atoms and C atoms also suddenly increases to above 0.7, but its distribution is a square-like ring instead of a circular ring, which indicates that maybe both a covalent bond and an ionic bond are formed between the Mn atom and the C atom.

Fig. 5 (g) shows the ELF profiles along the path from the Mn3c site to the X site and then to the Mn3c site as indicated by the black arrows in Fig. 5 (a-c) to further confirm the chemical bond type in $Mn_4X$. In the case of vacancy, the ELF value is stable at 0.35 over a wide range of interactions between the two Mn3c sites, accurately demonstrating the itinerant electrons between Mn atoms in the metallic bond. In the case of N interstitial, the ELF value reaches 0.8 only in a very narrow range between the Mn atom and the N atom close to the N atom side, which intuitively shows the position of the localized shared electron pair between the Mn atom and the N atom in the covalent bond. The situation of C interstitial is similar to that of N interstitial, except that the ELF value drops to 0.01 at the center of C atom site, reflecting the delocalization of electrons.

Using the ELF method, the type of chemical bonds is confirmed in $Mn_4X$. Next, the charge density difference (CDD) method is utilized to investigate the spatial distribution of chemical bonds and further determine the orbital composition involved in bonding.

In order to reflect the changes in the charge density of each component atom during the chemical bonding of $Mn_4X$, we performed the following CDD calculation:

$$CDD = CD_{Mn4X} - CD_{Mn1} - CD_{Mn2} - CD_{Mn3} - CD_{Mn4} - CD_X$$

Where $CD_{Mn4X}$, $CD_{Mn1}$, $CD_{Mn2}$, $CD_{Mn3}$, $CD_{Mn4}$ and $CD_X$ represent the charge density when the crystal structure of Figure. 2 (b) is fully occupied by four Mn atoms and one X atom, occupied only by one Mn atom at the sublattice 1 site, occupied only by one Mn atom at the sublattice 2 site, occupied only by one Mn atom at the sublattice 3 site, occupied only by one Mn atom at the sublattice 4 site, and occupied only by one X atom at position X, respectively.

Figure. 6 (a-c) shows the CDD mapping on the pink (001) lattice plane shown in Figure. 4, and Figure. 6 (d-f) shows the CDD mapping on the blue (001) lattice plane shown in Figure. 4. From left to right are the cases of vacancy interstitials, N interstitials, and C interstitials, respectively. It is worth noting that Figure. 6 (a-f) and Figure. 5 (a-f) correspond to the same two-dimensional lattice plane and the same interstitial situation, which facilitates the analysis of chemical bonding. In addition, the CDD scale in Fig. 6 (d-f) is moderately reduced to between -0.03 to 0.01, which is smaller than that in Fig. 6 (a-c), to enhance the contrast. In Fig. 6 (a-f), the black contour lines correspond to the distribution of lost electrons, and the white contour lines correspond to the distribution of gained electrons. This makes it easier for us to determine which electron orbitals contribute to the above distributions. Electron orbitals have strong anisotropy. From Figure. 4, it can be seen that the electron orbitals can be distinguished and determined by their orientation relative to the coordinate axis. For example, the $e_g$ state of Mn atoms is composed of orbitals along the coordinate axis, including $d_{3x^2-r^2}$ and $d_{y^2-z^2}$ orbitals, while the $t_{2g}$ state of Mn atoms consists of orbitals along the direction that is 45 degrees to the coordinate axis, including $d_{xy}$, $d_{yz}$ and $d_{zx}$ orbitals, and the 2p orbitals of X atoms are all along the coordinate axis, including $p_x$, $p_y$ and $p_z$ orbitals. Therefore, we also plotted the $x$ and $y$ coordinate axes in Figure. 6 in order to facilitate identification of which electron orbitals participate in bonding in Mn$_4$X.

There are three characteristics on the value and distribution of CDD in Fig. 6 (a-i). First, under all different interstitial settings the distribution of CDD between the Mn1a site and the Mn3c site and between the Mn3c site and the Mn3c site in Fig. 6 (a-f) are almost the same. Specifically, there is a positive CDD between the nearest neighboring Mn atoms, and its direction is 45 degrees to the coordinate axes, showing strong anisotropy. This is not a characteristic of metallic bonds but a characteristic of covalent bonds, which corresponds exactly to the black elliptical contours in the ELF of Figure. 5 (a, d-f). Furthermore, the above distribution of CDD shows that covalent bonding occurs in the $d_{xy}$ orbitals between neighboring Mn atoms as indicated by the electronic orbital bonding diagram in Figure. 6 (a). Whether from the perspective of energy or orbital direction, the covalency of $d_{xy}$ orbitals between neighboring Mn atoms is allowed. Fig. 6 (g) and Fig. 6 (h) show the CDD profiles along the path from the Mn3c site to the Mn1a site and from the Mn3c site to the Mn3c site as indicated by the black and red arrows in Fig. 6 (e) and Fig. 6 (b), respectively, to reflect the covalent bond strength in Mn$_4$X. It can be seen that both the N interstitial and the C interstitial would enhance the $d_{xy}$ orbital covalency between Mn1a site and Mn3c site through increasing the localization of electrons on the $d_{xy}$ orbital of the Mn1a site, but would suppress the $d_{xy}$ orbital covalency between Mn3c site and Mn3c site by decreasing the localization of electrons on the $d_{xy}$ orbital of the Mn3c site. This situation is particularly obvious for N interstitial, which may be related to the stronger electronegativity of N.

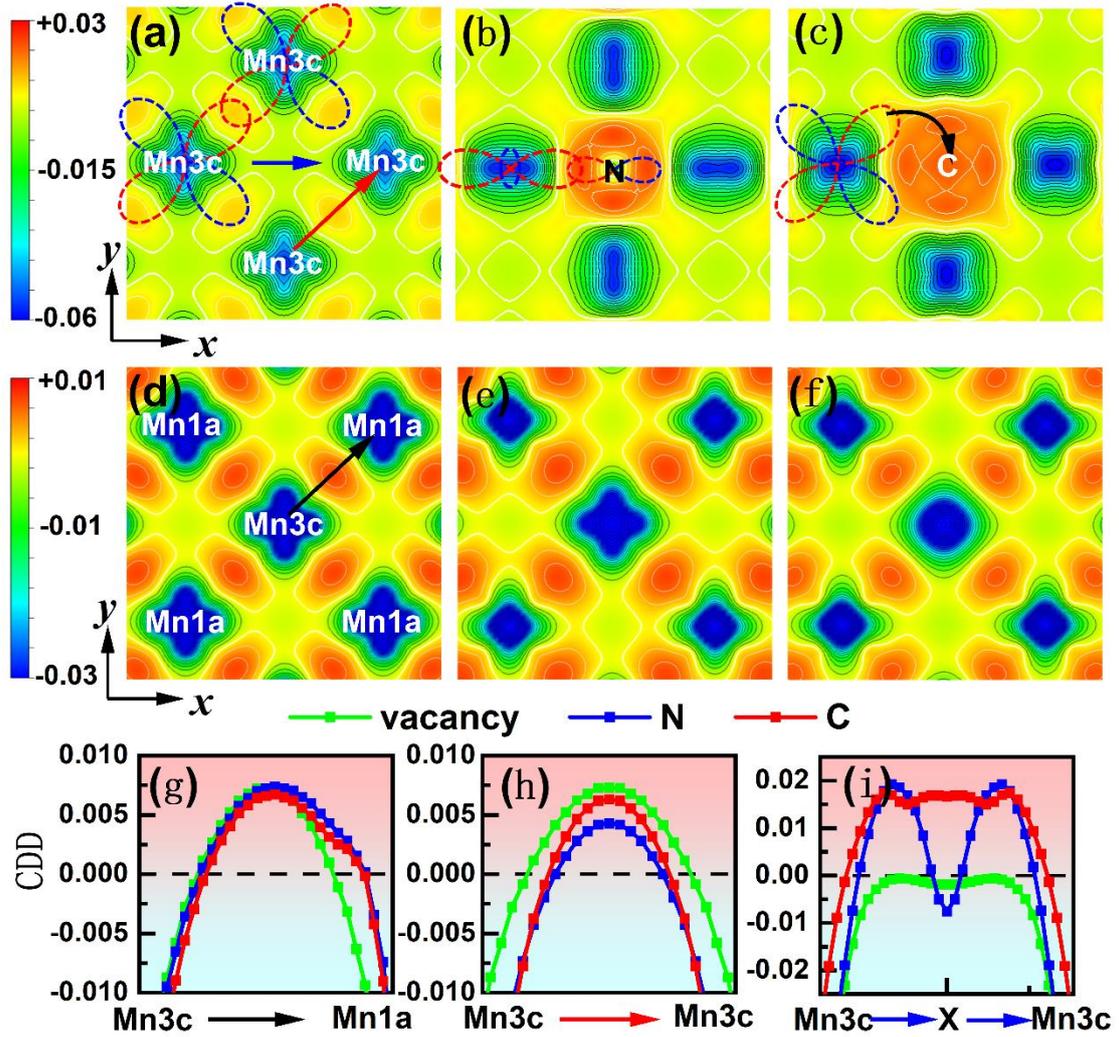

Figure. 6. Strength and spatial distribution of chemical bonds in $Mn_4X$. Charge density difference (CDD) mapping on the pink (001) lattice plane shown in Figure. 4 under vacancy interstitial (a), N interstitial (b), and C interstitial (c). CDD mapping on the blue (001) lattice plane shown in Figure. 4 under vacancy interstitial (d), N interstitial (e), and C interstitial (f). The white and black contour lines of CDD in (a-f) are positive and negative, respectively. The blue and red dashed lines are schematic electron orbitals. (g) CDD profiles along the path from Mn3c site to Mn1a site as indicated by the black arrow in (d). (h) CDD profiles along the path from Mn3c site to Mn3c site as indicated by the red arrow in (a). (i) CDD profiles along the path from Mn3c site to X site and then to Mn3c site as indicated by the blue arrow in (a).

Second, in the case of N interstitials as shown in Fig. 6 (b), an obvious positive CDD along two coordinate axes suddenly appears between Mn3c site and N and is very close to the N atoms as indicated by the white contour lines in the red area, and an obvious negative CDD of the Mn atom only along one coordinate axis also appears at the same time as shown by the black contour line in the blue area. Thus, the CDD distribution is strongly anisotropic, and meets the characteristics of the $d_{3x^2-r^2}$ or $d_{y^2-z^2}$ orbital of Mn atoms and the $p_x$ and $p_y$ orbitals of N atoms. That is to say, the electrons in the the $d_{3x^2-r^2}$ or $d_{y^2-z^2}$ orbitals of Mn atoms form polar covalent bonds with the electrons in the $p_x$ and $p_y$ orbitals of N atoms. The orbital cartoon in Figure. 6 (b) illustrates the covalent bond between the $d_{3x^2-r^2}$ orbital of Mn and the $p_x$ orbital of N. The CDD profile along the path from Mn3c site to N site and then to Mn3c site (the blue curve) in Fig. 6 (i) also reflects the polar covalent bonds more intuitively from a numerical level. Fig. 6 (b) and (i) show that the CDD at the center of the N atom is negative, rather than positive, which means that no electron is transferred from Mn to N despite the electronegativity of N. This may be related to the fact that the $p_x$, $p_y$ and $p_z$ orbitals of the N atom are occupied by exactly one electron. Electron diffraction [44] and X-ray diffraction experiments [45] also confirmed that there is only polar covalent bond but no ionic bond between Mn and N. Due to the strong polar covalent bonding, the $d_{3x^2-r^2}$ and $d_{y^2-z^2}$ orbitals of the adjacent Mn atoms can overlap directly without passing through the $p_x$, $p_y$ and $p_z$ orbitals of N atoms.

Third, for the case of C interstitials as shown in Fig. 6 (c) and (i), it is clear that covalent bonding also occurs between the $p_x$ and $p_y$ orbitals of C atoms and the $d_{3x^2-r^2}$ or $d_{y^2-z^2}$ orbitals of Mn atoms, but compared with the case of N interstitial, the shared electron pair is not so close to the C atom and the positive value of CDD is also smaller. This will greatly reduce the probability of direct overlap between the $d_{3x^2-r^2}$ and $d_{y^2-z^2}$ orbitals of the adjacent Mn atoms. Another very distinct feature of the C interstitial is that the distribution of negative CDD at the Mn3c site is not only along the coordinate axis but also along the direction 45 degrees to the coordinate axes, and there is an obvious positive CDD distribution just at the center of the C site which is different from the negative CDD distribution at the center of the N site. The electronegativity of C is weaker than that of N, but only two electrons in the isolated C atom occupy the $p$ orbital, resulting in one of the $p$ suborbitals being completely vacant, thus providing the necessary conditions for electron transfer. Judging from the direction and energy characteristics of the $d_{xy}$ orbital shown in Figure. 4, the electron transfer should be from the $d_{xy}$ orbital of Mn to the $p$ orbital of C. The orbital cartoon in Figure. 6 (c) illustrates this kind of electron transfer. Combined with the strong delocalization of the center of the C site shown in Fig. 5 (c) and (g), the the $p_x$, $p_y$ and $p_z$ orbitals orbitals of C atoms can serve as media for the interaction between the the $d_{3x^2-r^2}$ and $d_{y^2-z^2}$ orbitals of neighboring Mn atoms.

In addition, by comparing the CDD values in Figure. 6 (g-i), it can be found that the $d$-$d$ covalent bonding between Mn and Mn is much weaker than the $p$-$d$ covalent bonding between Mn and N or C. That is to say, the degree of orbital overlap in the system is greatly enhanced after N or C interstitials.

## C. Manipulation of magnetic structure by chemical bonds

Electrons are the binding agents of materials, and chemical bonds are the direct manifestation of this bind effect, which determines all the physical properties of the material, including magnetism. Much research has centered on the role of chemical bonds in stabilizing crystal structures, while this discussion will explore their impact on stabilizing magnetic configurations. The magnetic configuration of the system is governed by the spin-dependent Hamiltonian, primarily encompassing the Heisenberg exchange interaction $\sum_{ij} J_{ij} S_i \cdot S_j$, magnetocrystalline anisotropy $\sum_i K_i (S_i \cdot e_i)^2$, and Dzyaloshinskii-Moriya interaction $\sum_{ij} D_{ij} \cdot (S_i \times S_j)$. In centrosymmetric Mn$_4$X, the Dzyaloshinskii-Moriya interaction is absent.[46] The Heisenberg exchange interaction and magnetocrystalline anisotropy are the principal interactions, which are dictated by the extent and manner of electron cloud overlap within covalent bonds. In the previous discussion of this paper, we clarified the ground state magnetic structure and exchange integral constant of the Mn triangular lattice magnet, and also obtained parameters such as the type, strength and spatial distribution of chemical bonds. Now let us analyze in detail how chemical bonds affect the selection of the magnetic configuration of Mn triangular lattice magnets.

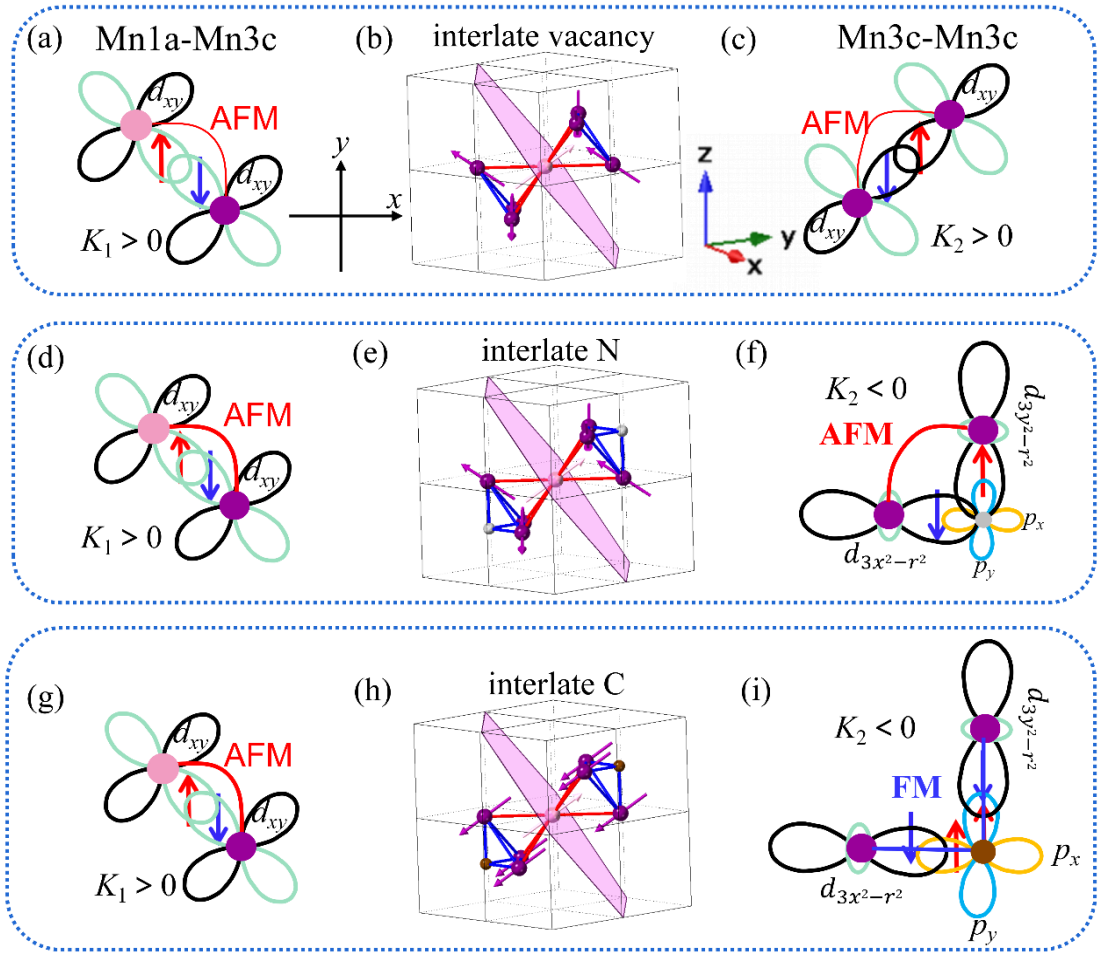

Figure. 7. Relationship between orbital bonding and magnetic configuration in $Mn_4X$. Pink and purple spheres represent Mn atoms occupying Mn1a and Mn3c sites, respectively. Gray and brown spheres represent N and C atoms occupying X sites, respectively. Schematic diagram of orbital bonding and exchange interactions between Mn1a and Mn3c sites under vacancy interstitial (a), N interstitial (d), and C interstitial (g). Schematic diagram of orbital bonding and exchange interactions between Mn3c and Mn3c sites under vacancy interstitial (c), N interstitial (f), and C interstitial (i). AFM and FM represent antiferromagnetic exchange interaction and ferromagnetic exchange interaction, respectively. The thickness of the solid lines reflects the strength of the exchange interaction. Schematic diagram of magnetic configurations under vacancy interstitial (b), N interstitial (e), and C interstitial (h).

Let us first examine the simplest case of vacancy interstitial. As illustrated in

Figure. 4, the cubic crystal field lifts up the $t_{2g}$ orbital. Consequently, the exchange interactions both between Mn1a and Mn3c and between Mn3c and Mn3c are mediated by the covalent bonding of $t_{2g}$ orbitals, as depicted in Figure. 6 (a) (d), this direct overlap between the *d-d* orbital wave functions provides a direct exchange interaction. On the one hand, according to Pauli's exclusion principle, the *d-d* direct exchange interaction is AFM, i. e., $J_1 > 0$ and $J_2 > 0$, as shown in Figure. 7 (a) (c), which is consistent with the results in Figure. 3 (b). On the other hand, the anisotropy of the $t_{2g}$ orbitals in Mn atoms within cubic crystals induces a positive magnetocrystalline anisotropy $K_1 > 0$ and $K_2 > 0$, corresponding to easy-plane anisotropy [47]. In this scenario, solely considering the interactions $J_2 > 0$ and $K > 0$ among Mn3c sites results in a $\Gamma_{5g}$ coplanar magnetic structure [47]. Given that Mn1a and Mn3c sites form a perfect tetrahedron, introducing $J_1 > 0$ interactions among Mn1a and Mn3c sites distorts the $\Gamma_{5g}$ structure into an all-in-all-out non-coplanar magnetic configuration. In this arrangement, each Mn atom's spin direction points to the tetrahedron's center, that is, $\theta = \alpha = 109.5°$, as illustrated in Figure. 7 (b), which is exactly the magnetic ground state shown in Figure. 2 (d). Therefore, the chiral spin state in $\gamma$-Mn is manipulated by the covalent bonding of *d* orbitals in the $t_{2g}$ state.

Next, let us consider the case of N interstitial. The chemical bonding at the Mn1a site remains unchanged, continuing to form covalent bonds with Mn3c via $t_{2g}$ orbitals, thus maintaining $J_1 > 0$ and $K_1 > 0$ [47,48]. However, the bond strength has increased due to enhanced *d-d* electron cloud overlap, as shown in Figure. 6 (g), significantly raising AFM direct exchange interaction, as shown in Figure. 7 (d). The chemical bonds at the Mn3c site have undergone significant changes. Firstly, the *d-d* covalency between Mn3c atoms is markedly weakened, as shown in Figure. 6 (h), suppressing the AFM direct exchange interaction between their $t_{2g}$ orbitals. Secondly, Mn3c forms a strong *p-d* covalent bond with the N atom via the $e_g$ orbital, as illustrated in Figure. 6 (b) (i), transforming the magnetocrystalline anisotropy of Mn3c from easy-plane ($K_2 > 0$) to easy-axis ($K_2 < 0$) [47]. More importantly, the strong electronegativity of the N atom enhances electron cloud overlap among the $e_g$ orbitals of Mn3c, creating a new AFM direct exchange interaction ($J_2 > 0$) channel, as shown in Figure. 7 (f). Consequently, the N interstitial enhances the AFM exchange interaction at both Mn1a and Mn3c sites and introduces easy-axis anisotropy at Mn3c site, all of which will make the all-in-all-out magnetic ground state in Mn$_4$N more stable than that in $\gamma$-Mn, as shown in Figure. 7 (e). The insights regarding $J_1$, $J_2$ and the magnetic ground state derived from the analysis of chemical bonds are fully consistent with the results obtained through first-principles calculations shown in Figure. 2 (e) and 3 (b). This consistency underscores that chemical bonding is fundamental to the formation of magnetic configurations.

Finally, let us discuss the case of C interstitial. The chemical bonding at the Mn1a site, along with the resulting $J_1$ and $K_1$ values, is identical to that observed with the N interstitial. Figure. 7 (g) illustrates this case. At the Mn3c site, the $d$-$d$ covalency via the $t_{2g}$ orbital is weakened, and a strong $p$-$d$ covalency with C through the $e_g$ orbital is formed, similar to the case with N interstitials. However, unlike N, the C atom has one completely empty $p$ orbital among three $p$ orbitals, leading to the transfer of electrons from the $t_{2g}$ orbitals of Mn3c to the empty $p$ orbital. These transferred electrons are strongly delocalized, as shown in Figure. 5 (c) (g) and Figure. 6 (c) (i), which provide a medium for super exchange interactions [49]. As a result, a 90º FM super exchange interaction between Mn3c-C-Mn3c is formed [50-55], as shown in Figure. 7 (i). Since the electronegativity of C is weaker than that of N, the $p$-$d$ bonding electrons are not as concentrated on C. This results in the absence of an AFM direct exchange interaction between the $e_g$ orbitals of Mn3c. The analysis of the chemical bonds in Mn$_4$C elucidates the origin of the $J_2 < 0$ value obtained from our first-principles calculations as shown in Figure. 3 (b). Therefore, the C interstitial enhances the AFM direct exchange interaction at the Mn1a sites and introduces a FM super exchange interaction at the Mn3c sites [55,56]. It also introduces easy-axis anisotropy ($K_2 < 0$) at the Mn3c site. These factors collectively form the basis of the 3:1 collinear magnetic ground state in Mn$_4$C, as shown in Figure. 7 (h).

## IV. CONCLUSION

In summary, our aim is to understand and manipulate the internal driving forces for the formation of chiral spin states in Mn triangular lattice magnets from the perspective of chemical bonds. To this end, on the one hand, it is found that both the vacancy interstitial in $\gamma$-Mn and the N interstitial in $Mn_4N$ stabilize the all-in-all-out chiral spin magnetic ground state, while the C interstitial in $Mn_4C$ stabilize the 3:1 collinear ferrimagnetic ground state. On the other hand, electron localization function (ELF) and charge density difference (CDD) analyses revealed that the type, strength and spatial distribution of chemical bonds in Mn triangular lattice magnets were regulated by the X (X = vacancy, N and C) interstitials. On this basis, we found that through the magnetic exchange integral constant and magnetocrystalline anisotropy the weak nonpolar covalent bonds between $t_{2g}$ orbitals of Mn under vacancy interstitial construct the all-in-all-out chiral spin state, compared with vacancy interstitials, under N interstitials both the newly introduced strong polar covalent bond between the $e_g$ orbital of Mn3c and the $p$ orbital of N and the enhanced covalent bond between the $t_{2g}$ orbital of Mn1a and the $t_{2g}$ orbital of Mn3c make the chiral spin state more stable, in contrast, under C interstitials the ionic bond between the $t_{2g}$ orbital of Mn3c and the $p$ orbital of C greatly stabilizes the 3:1 collinear ferrimagnetic structure non-chiral spin state. Our work points out that the chiral spin states can be regulated by regulating the chemical bonds in Mn triangular lattice magnets through different interstitial light elements.

## Acknowledgements

This work was supported by the National Natural Science Foundation of China (grant no. 51901067, 51971087, 52101233, and 52071279), the Natural Science Foundation of Hebei Province (grant no. E2019205234), the Science and Technology Research Project of Hebei Higher Education (grant no. QN2019154), and the Science Foundation of Hebei Normal University (grant no. L2019B11), and the "333 Talent Project" of Hebei province (Grant No. C20231105) and the Science Foundation of Hebei Normal University, China (Grant No. L2024B08).